\documentclass{article}

\usepackage{microtype}
\usepackage{graphicx}
\usepackage{booktabs} 

\usepackage{hyperref}

\usepackage[accepted]{icml2019}

\icmltitlerunning{On the Long-term Impact of Algorithmic Decision Policies}

\usepackage{amsmath}
\usepackage{natbib}
\usepackage{color}
\usepackage{amsfonts}
\usepackage{amssymb}
\usepackage{graphicx}

\newtheorem{definition}{Definition}

\newcommand{\argmin}{\arg \min}

\newcommand{\Exp}{\mathbb{E}}

\newcommand{\reals}{\mathbb{R}}

\newcommand{\vx}{\mathbf{x}}

\newcommand{\vz}{\mathbf{z}}

\newcommand{\hy}{\hat{y}}
\newcommand{\hY}{\hat{Y}}

\newcommand{\cH}{\mathcal{H}}

\newcommand{\cY}{\mathcal{Y}}
\newcommand{\cX}{\mathcal{X}}
\newcommand{\cZ}{\mathcal{Z}}

\newcommand{\eff}{\mathcal{E}}
\newcommand{\rew}{\mathcal{R}}
\newcommand{\util}{\mathcal{U}}

\newcommand{\hh}[1]{\textcolor{black}{#1}}
\newcommand{\modif}[1]{\textcolor{black}{#1}}

\usepackage{subcaption}

\begin{document}

\twocolumn[
\icmltitle{On the Long-term Impact of Algorithmic Decision Policies:\\ Effort Unfairness and Feature Segregation through Social Learning}



\icmlsetsymbol{equal}{*}

\begin{icmlauthorlist}
\icmlauthor{Hoda Heidari}{equal,ethz}
\icmlauthor{Vedant Nanda}{equal,mpi}
\icmlauthor{Krishna P. Gummadi}{mpi}
\end{icmlauthorlist}

\icmlaffiliation{ethz}{Computer Science Department, ETH Z{\"u}rich, Z{\"u}rich, Switzerland}
\icmlaffiliation{mpi}{MPI-SWS, Saarbr{\"u}cken, Germany}

\icmlcorrespondingauthor{Hoda Heidari}{hheidari@inf.ethz.ch}


\icmlkeywords{Fairness, Effort, Long-term, Impact}

\vskip 0.3in
]



\printAffiliationsAndNotice{\icmlEqualContribution} 

\begin{abstract}
Most existing notions of algorithmic fairness are \emph{one-shot}: they ensure some form of allocative equality at the time of decision making, but do not account for the \emph{adverse impact} of the algorithmic decisions today on the \emph{long-term} welfare and prosperity of certain segments of the population. 
We take a broader perspective on algorithmic fairness. We propose an \emph{effort}-based measure of fairness and present a data-driven framework for characterizing the long-term impact of algorithmic policies on reshaping the underlying population. Motivated by the psychological literature on \emph{social learning} and the economic literature on \emph{equality of opportunity}, we propose a micro-scale model of how individuals may respond to decision making algorithms. We employ existing measures of segregation from sociology and economics to quantify the resulting macro-scale population-level change. Importantly, we observe that different models may shift the group-conditional distribution of qualifications in  different directions. Our findings raise a number of important questions regarding the formalization of fairness for decision-making models.
\end{abstract}
\section{Introduction}

Machine Learning tools are increasingly employed to make consequential decisions for human subjects, in areas such as credit lending~\citep{whitecase}, policing~\citep{policing}, criminal justice~\citep{sentencing}, and medicine~\citep{deo2015machine}. Decisions made by these algorithms can have a long-lasting impact on people's lives and may affect certain individuals or social groups negatively~\citep{sweeney2013discrimination,propublica,guardian_beauty}. This realization has recently spawned an active area of research into quantifying and guaranteeing fairness for machine learning~\citep{dwork2012fairness,kleinberg2016inherent,hardt2016equality}. 

Most existing notions of fairness assume a \emph{static} population: they ensure some form of \emph{allocative equality} at the time of decision making, but do not account for the \emph{adverse impact} of algorithmic decisions today on the \emph{long term} welfare and prosperity of different segments of the population. For instance, consider equality of odds~\citep{hardt2016equality}. The notion requires that the model distributes different types of error (i.e., false positives and false negatives) equally across different social groups. But it does not take into consideration the fact that for members of the advantaged group these erroneous predictions may be easy to overturn, whereas for the disadvantaged it may take a significant amount of \emph{effort} to improve their qualifications to obtain better algorithmic outcomes. Furthermore, in the long run, the decision-making model may nudge different segments of the population to obtain very different sets of qualifications---some of which might be socially and economically more desirable than others. This may in effect lead to further \emph{marginalization} of these groups. 

Motivated by these concerns about existing notions of fairness, we argue for a broader view of algorithmic models---one that treats them as \emph{policies} implemented within a social context and with the potential of impacting individuals and reshaping society. Among other considerations\footnote{Another important factor is how a utility maximizing decision maker---employing the model---would respond to its predictions. For instance, they may interpret the predictions in a certain way, or update the model entirely. Prior work~\citep{liu2018delayed,kannan2018downstream} has already addressed some of these considerations.}, such view of decision-making models necessitates a deeper understanding of how individual decision subjects may \emph{respond} to these models and how those responses may translate into \emph{adverse impact} for certain segments of the population. 

In this work, we propose an effort-based measure of unfairness for algorithmic decisions.
We define a data-driven, group-dependent measure of \emph{effort} drawing on the economic literature on Equality of Opportunity~\citep{roemer2015equality}. Our effort function captures the idea that the kind of changes required to obtain a desirable algorithmic outcome (e.g., changing one's school type from public to private to get a better prediction for SAT score) is often significantly more difficult to make for members of the disadvantaged group compared to the advantaged. Building on this notion of effort, we formulate \emph{effort unfairness} as the inequality in the amount of effort required for members of each group to obtain their desired outcomes. 

To formulate the long-term impact of algorithmic policies on the underlying population, we specify a micro-scale model of how individuals respond to algorithmic decision-making models, taking inspiration from the psychological literature on \emph{social learning}~\citep{bandura1962social,bandura1978social}. We posit that individuals observe and imitate the qualifications of their \emph{social models}---someone who has received a better algorithmic outcome from the decision-making model---if by doing so, they can obtain higher rewards~\citep{bandura1962social,apesteguia2007imitation}. More precisely, we model an individual's response to be the decision-making algorithm by first selecting a social model for him/her; the individual is then assumed to exert effort to attain his/her model's qualifications if and only if doing so improves his/her overall utility. 
With this individual-level behavioral model in place, we can simulate decision subjects' responses and quantify the macro-scale impact of algorithmic policies on reshaping the underlying populations. We employ existing measures of segregation from sociology and economics~\citep{massey1988dimensions} to characterize how the distribution of qualification for each group changes in response to the deployed model.  Importantly, we observe that different models may shift the group-conditional distribution of qualifications in vastly different directions. 

Our work raises a number of important questions about algorithmic policies and the formulation of fairness: What is the ultimate purpose of a fair predictive model---to guarantee allocative equality today, or to ensure similar distributions of qualifications in the long-run? With respect to short term allocative equality, are all error created equal, or should we take into account the disparity in the effort it takes for different groups to obtain their desired predictions? In the long run, \hh{what are the type of changes that different predictive models impose on the society? Which ones are desirable, and which ones should be watched out for?} Is it ethically and economically acceptable to nudge different segments of the population toward obtaining different qualifications? If not, how can we prevent this without employing a model whose decisions may be perceived as unfair today?
 These are all critical questions that must be carefully analyzed before determining which model is fair and best-suited to make consequential decisions for humans. Addressing such ethical challenges is outside the scope of this paper---and arguably intradisciplinary Machine Learning research. We hope that our work serves as a reminder to the ML community that to formalize fairness appropriately we need to first formalize the processes and dynamics through which algorithmic decisions impact their subjects and society in the long run. 

\subsection{Related Work}\label{sec:related}
Most existing notions of algorithmic fairness are one-shot and require that a particular error metric is equal across all social groups. Different choices for the metric have led to different fairness criteria; examples include demographic parity~\citep{kleinberg2016inherent,dwork2012fairness,corbett2017algorithmic}, disparate impact~\citep{zafar2017fairness,feldman2015certifying}, equality of odds~\citep{hardt2016equality}, and calibration~\citep{kleinberg2016inherent}. Prior notions fail to capture the disparity in the \emph{effort} it takes members of different social groups to improve their algorithmic outcome. 
%
We propose a group-dependent, data-driven measure of effort, inspired by the literature on Equality Of Opportunity (EOP)~\citep{roemer2015equality,heidari2019a}. (In particular, the effort it takes individual $i$ to improve their feature $k$ value from $x$ to $x'$ is proportional to the difference between the rank/quantile of $x'$ and $x$ in the distribution of feature $k$ in $i$'s social group.)

Social (or observational) learning~\citep{bandura2008observational} is a type of learning that occurs through observing and imitating the behavior of others. This type of learning requires a social model (or role model)---someone of higher status in the environment. According to the social learning theory, observers recreate their role model's behavior only if they have sufficient motivation~\citep{bandura1962social,apesteguia2007imitation}---this often comes from the observation that the model is rewarded for their actions. In our model, an individual recreates their role model's qualification if by doing so he obtains a positive utility, where utility is defined as reward minus effort. 
Furthermore, it has been shown that observers learn best from models that they identify with\footnote{Social identity is a person's sense of who they are based on their group membership(s)~\citep{tajfel1979integrative}.} and find it within their capability to imitate them~\citep{bandura1962social}. These points are captured by our effort function.  \modif{Social learning explicitly captures the role model implications of decision making policies. This echos research in sociology and economics which has already established the role model effects of affirmative action policies~\citep{chung2000role}. }We note that imitation dynamics have been extensively studied in population and evolutionary games (see, e.g.,~\citep{sandholm2010population}, Chapters 4 and 5). 
%

Several recent papers study the impact of decision-making models and fairness interventions on society and individuals (see, e.g.,~\citep{liu2018delayed,kannan2018downstream}). Unlike prior work, our focus is on \emph{how subjects respond} to algorithmic policies by \emph{improving/updating their (mutable) qualifications}. We don't make any case-specific assumptions about how the world changes in response to the deployed model, rather allow our micro-scale behavioral model to derive the macro-level change. We emphasize that our model is not meant to perfectly capture all the behavioral nuances involved; rather our primary goal is to highlight the potential role of behavioral dynamics and human responses in shaping the long-term impact of algorithmic models.

Also related but orthogonal to our work is a recent line of research on \emph{strategic classification}---a setting in which decision subjects are assumed to respond \emph{strategically} and potentially \emph{untruthfully} to the choice of the classification model, and the goal is to design classifiers that are robust to strategic manipulation~\citep{dong2018strategic,hu2018disparate,milli2018social}. 

\section{Setting}\label{sec:setting}


We consider the standard supervised learning setting: A learning algorithm receives the training data set $D=\{(\vx_i,y_i)\}_{i=1}^n$ consisting of $n$ instances, where $\vx_i \in \cX$ specifies the feature vector for individual $i$ and $y_i \in \cY$, the \textit{ground truth} label for him/her. We use $s_i$ to refer to the sensitive feature value (e.g., race, gender, or their intersection) for individual $i$. 
For the ease of notation, we will use $\vz_i$ to denote the example $(\vx_i,y_i)$. We assume $\vz_i$ fully characterizes individual $i$ with respect to the task at hand. 
The training data is sampled i.i.d. from a distribution $P$ on $\cZ=\cX \times\cY$. For simplicity, throughout we assume there exists an unknown function $f:\cX \rightarrow \cY$ such that for all $i$, $y_i = f(\vx_i)$. 
Unless specified otherwise, we assume $\cX \subseteq \reals^K$, where $K$ denotes the number of features. 
The goal of a learning algorithm is to use the training data to fit a (regression) \emph{model} (or hypothesis) $h: \cX \rightarrow \cY$ that accurately predicts the label for new instances. Let $\cH$ be the hypothesis class consisting of all the models available to the learning algorithm.
A learning algorithm receives $D$ as the input; then utilizes the data to select a model $h \in \cH$ that minimizes some notion of loss, $L$. For instance, in regression the empirical mean squared loss of a model $h$ on $D$ is defined as $L_D(h) = \sum_{i\in D}  ( y_i  - \hy_i )^2$,
where $\hy_i = h(\vx_i)$.
The learning algorithm outputs the model $h \in \cH$ that minimizes the empirical loss; i.e.,
$h = \argmin_{h'} L_D(h')$.

We assume there exists a benefit function $b: \cX, \cY \times \cY \rightarrow \reals$ that quantifies the benefit an individual with feature vector $\vx$ and ground truth label $y$ receives if the trained model predicts label $\hy$ for them. Throughout this work we will focus on benefit functions that are only functions of $y, \hy$ and are linear in $\hy$ (e.g., $b(y, \hy) = \hy - y + 1$ or $b(y, \hy) = \hy$). For simplicity and ease of interpretation, all illustrations in the main body of the paper are performed with $\hy$ as the benefit function. Throughout, we assume higher predicted labels are considered more desirable from the point of view of individual decision subjects (e.g., this is the case when the task is to predict students' grade to decide who is admitted to a top school).

\subsection{Efforts, Rewards, Utilities}
Let $h$ specify the deployed predictive model. Consider an individual characterized by $\vz = (\vx,y)$. 
Let $\rew_h(\vz, \vz')$ specify the \emph{reward} or added benefit he/she obtains as the result of changing his/her characteristics from $\vz$ to $\vz'$:
$$\rew_h(\vz, \vz') = b(h(\vx'), y')^\alpha - b(h(\vx), y)^\alpha,$$
where $\alpha>0$ is a constant specifying the individual's degree of risk aversion. This parameter can be adjusted to model diminishing returns to added benefit. Unless otherwise specified, in our illustrations we take $\alpha=1$.
Let $\eff_h(\vz, \vz')$ specify the effort it takes the individual to update their qualifications and make the change from $\vz$ to $\vz'$ (we will shortly elaborate on how effort can be quantified). The overall utility of the individual is denoted by $\util_h(\vz, \vz')$ and for simplicity, we assume it takes on a a linear form: 
$$\util_h(\vz,\vz') = \rew_h(\vz,\vz') - \eff_h(\vz, \vz').$$
That is, utility is simply reward minus effort.
When clear from the context, we drop the subscript $h$.

Throughout, we focus on effort functions that only depend on $\vx,\vx', s$. For simplicity, we assume the effort function is additively separable across features---that is, the total effort required to change $\vx$ to $\vx'$ is a linear combination of the effort needed to change each feature separately:\footnote{Depending on our domain knowledge of how features relate to one another, we may find a different aggregating operator (e.g., max) to be more appropriate than summation. For instance, if two features automatically change together, no extra effort is required for changing both of them simultaneously. Throughout our illustrations, for simplicity we focus on additively separable functions, but our results can be readily produced for more complicated effort functions.}
\begin{equation*}\label{eq:eff}
\eff(\vz, \vz') = \eff_s(\vx, \vx') = c_s + \frac{1}{K}\sum_{k=1}^K c_{s,k} \epsilon_{s,k}(x_k,x'_k,),
\end{equation*}
where $\epsilon_{s,k}(x_k,x'_k,)$ denotes the effort it takes to change the value of feature $k$ from $x_k$ to $x'_k$ for an individual belonging to group $s$ (defined below). For group $s$, $c_s$ is a group-dependent constant specifying the minimum effort required to make any change. Similarly for $k=1,\cdots, K$, $c_{s,k}$'s are constant weights which allow us to specify the relative difficulty of change across different features. For simplicity, throughout we assume $c_{s,k} \equiv 1$ for all $k,s$. 

We define $\epsilon_{s,k}(x_k,x'_k,)$ as follows---depending on the feature type and our domain knowledge about the feature:\footnote{See the Appendix for a complete description of the effort function.}
	Suppose feature $k$ is numerical and \emph{monotone}~\citep{duivesteijn2008nearest}, that is, we expect an increase in its value to monotonically increase the predicted label---everything else being equal. Monotonicity implies that there is a clear direction of change that is considered desirable. As an example in the education context, consider the \emph{number of hours of study}: we expect an increase in this feature to increase the an student's predicted grade.
Without loss of generality, we assume higher values of feature $k$ are expected to increase the predicted label (we can ensure this by preprocessing the data and negating feature $k$ values if necessary). We define $\epsilon_{s,k}(x_k,x'_k,)$ as follows:
	$$\epsilon_{s,k}(x_k,x'_k,) = \max\{0, Q_{s,k}(x'_k) - Q_{s,k}(x_k)\}.$$
	Above, $Q_{s,k}(x)$ specifies the quantile/rank of value $x$ in the empirical distribution of feature $k$ among individuals who belong to group $s$. 
	
	The above effort function is inspired by the line of work on equality of opportunity~\citep{roemer2015equality}: Note that the distribution of feature $k$ values can be different across different social groups (e.g., men and women, or African-Americans and Whites). We take the view that this is potentially \emph{not} the result of one group being inherently inferior to another (in terms of feature $k$). Rather, it is most likely due to the underlying socio-economic circumstances that the privileged group can achieve higher values of feature $k$ with less effort. To account for this in our effort function, we measure the effort it takes a person $i$ in group $s$ to change their feature $k$ value from $x_k$ to $x'_k$ by comparing the rank/quantile of $x'_k$ and $x_k$ in the empirical distribution of feature $k$ within group $s$. This implies that if for most people in group $s$, the value of feature $k$ is equal or better than $x'_k$, we consider it relatively easy for individual $i$ to make the change from $x_k$ to $x'_k$. If however, very few in group $s$ have ever been able to achieve $x'_k$, then it is considered very difficult for $i$ to make this change.

Using the effort function defined above, in Section~\ref{sec:fairness} we propose a new effort-based measure of algorithmic unfairness. In Section~\ref{sec:long-term}, we will utilize our effort function to compute individual utilities and subsequently specifying the social models. 

\section{Effort-based Measures of Fairness}\label{sec:fairness}

\modif{Existing formulations of fairness are concerned with how errors are distributed among various social groups, but they do not account for the fact that even if errors are distributed similarly, the \emph{effort} required to fix those errors and improve one's prediction may be significantly higher for the disadvantaged subpopulation. }
Building on our notion of effort introduced in Section~\ref{sec:setting}, in this section we formulate a new measure of algorithmic unfairness, called the \emph{effort unfairness}. At a high level, effort unfairness is the disparity in the average effort members of each group have to exert to obtain their desired outcomes---by imitating the appropriate role models.\footnote{\modif{Note that algorithmic unfairness has many different aspects. We introduce a new dimension along which algorithmic decisions disparately affect different subpopulation, but this is not to undermine the importance of all other dimensions of unfairness (such as error disparities). In particular, we do \emph{not} claim that if a model is fair according to our notion, then it cannot be unfair according to other criteria. }}

We propose three different formalizations of effort unfairness: \emph{bounded-effort} unfairness, \emph{threshold-reward} unfairness, and \emph{effort-reward} unfairness. \hh{Each notion corresponds to a distinct/salient way in which decision subjects may evaluate fairness and respond to their predictions.}

\begin{figure}[h!]
	\centering
	\includegraphics[width=0.27\textwidth]{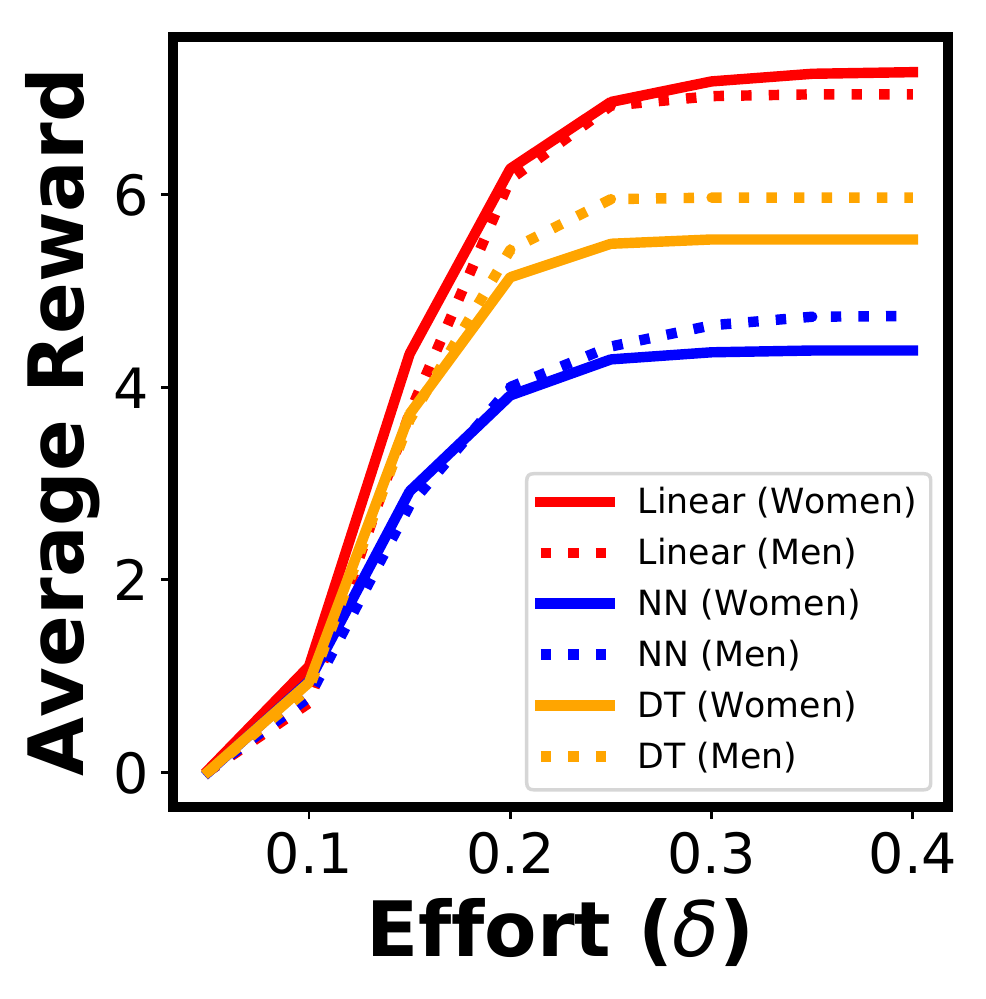}
	\caption{The average reward obtainable by members of each group as the result of exerting (at most) $\delta$ units of effort. For each model, the bounded effort unfairness is the difference between the dashed and solid curves. }
	\label{fig:bounded}
\end{figure}

\emph{Bounded-effort} unfairness is the inequality in the average reward members of each group can obtain by exerting a fixed level of effort. More precisely:
\begin{definition}[Bounded-effort Unfairness]
Given a constant $\delta$, the $\delta$-bounded-effort unfairness of a predictive model $h$ is the inequality of the following metric across different groups:
	\begin{equation}\label{eq:bounded}
	\Exp_{i \sim P: s_i =s} \left( \max_{\vz \in \cZ} \rew_h(\vz_i,\vz) \text{ s.t. } \eff_h(\vz_i,\vz) \leq \delta \right).
	\end{equation}
\end{definition}
The bounded-effort formulation is motivated by the literature on \emph{bounded willpower} in behavioral economics~\citep{mullainathan2000behavioral}, which at a high level posits that there is an upper bound on the level of effort people can be expected to exert. 

To compute the bounded-effort unfairness in practice, we propose replacing the expectation in Equation~\ref{eq:bounded} with the empirical mean, and taking the maximum over the available data set $D$. The latter not only simplifies the optimization, it also has a natural interpretation in terms of \emph{social models} (see Section~\ref{sec:long-term}):  Precisely, we estimate Equation~\ref{eq:bounded} by $	\frac{1}{n_s}\sum_{i: s_i =s} \left( \max_{\vz \in D} \rew_h(\vz_i,\vz) \text{ s.t. } \eff_h(\vz_i,\vz) \leq \delta \right)$ where $n_s$ is the number of subjects in group $s$.

Figure~\ref{fig:bounded} illustrates the bounded effort unfairness calculated over the student performance data set~\citep{cortez2008using}. (Information about the data set, our pre-processing steps, and the trained models can be found in the Appendix. The code used to generate the plots in this paper can be found at \url{https://github.com/nvedant07/effort_reward_fairness}.) Note that according to the bounded-effort measure, different models may be discriminatory against different groups. Also depending on the choice of $\delta$ the measure may rank the three models differently. 

\emph{Threshold-reward unfairness} is the inequality in the average effort member of each group need to exert to reach a certain level of reward. More precisely:
\begin{definition}[Threshold-reward Unfairness]
Given a constant $\delta$, the $\delta$-threshold-reward unfairness of a predictive model $h$ is the inequality of the following metric across different groups:
	\begin{equation}
	\Exp_{i \sim P: s_i =s}  \left( \min_{\vz \in \cZ} \eff_h(\vz_i,\vz) \text{ s.t. } \rew_h(\vz_i,\vz) \geq \delta \right).
	\end{equation}
\end{definition}
This formulation is motivated by the \emph{capability} view of fairness~\citep{sen1993capability}: Sen conceptualizes fairness as the equality of capability, where at a high level, capability is a person's ability to reach valuable states of being (in our case, a certain level of reward).

\begin{figure}[h!]
	\centering
	\includegraphics[width=0.3\textwidth]{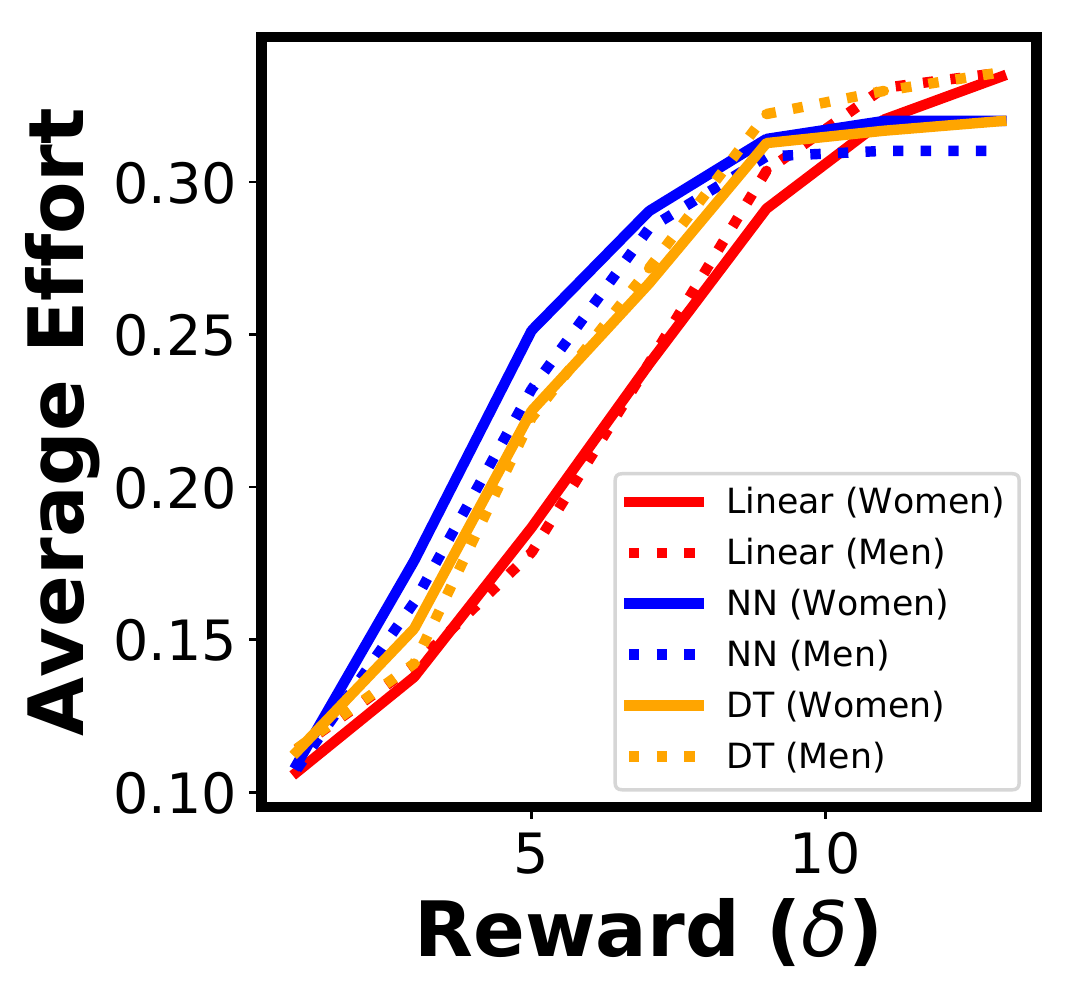}
	\caption{The average effort required from members of each group to obtain (at least) $\delta$ units of reward. For each model, the threshold-reward unfairness is the difference between the dashed and solid curves. }
	\label{fig:average}
\end{figure}

Figure~\ref{fig:average} illustrates the average effort unfairness on the student performance data set. Note that depending on the choice of $\delta$ the measure may rank the three models differently. Also interestingly, depending on the choice of $\delta$ the same model (i.e., linear model) may be considered unfair toward men, unfair toward women, or perfectly fair!

The previous two formulations of effort unfairness---while well-motivated---may give us different rankings across the same set of alternatives depending on our choice of $\delta$. The final formulation, which we call the \emph{effort-reward} unfairness, resolves this issue by comparing the highest utility members of each group can possibly achieve by exerting additional effort. More precisely:
\begin{definition}[Effort-reward Unfairness]
	For a predictive model $h$, the effort-reward unfairness is the inequality of the following metric across different groups:
	\begin{equation}
	\Exp_{i \sim P: s_i =s}  \max_{\vz \in \cZ} \util_h(\vz_i,\vz).
	\end{equation}
\end{definition}

\begin{figure}[h!]
	\centering
	\includegraphics[width=0.45\textwidth]{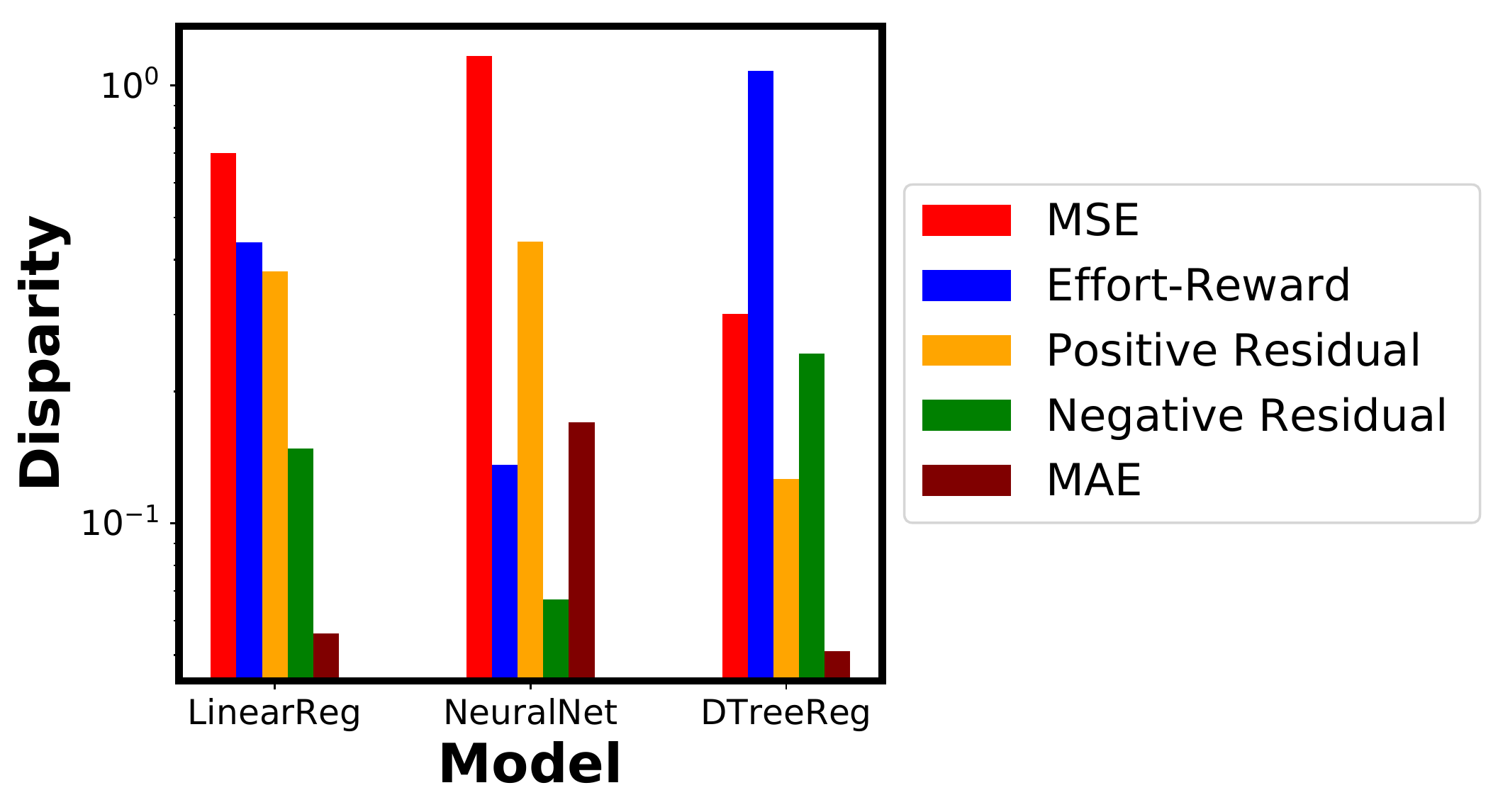}
	\caption{Comparison of the effort-reward unfairness with several existing notions of (un)fairness. Effort-reward unfairness ranks the models differently.}
	\label{fig:utility}
\end{figure}

Figure~\ref{fig:utility} contrasts the effort-reward measure with existing notions of algorithmic unfairness (these measures are precisely defined in the Appendix.) 
As evident in Figure~\ref{fig:utility}, on the student performance data set none of the existing fairness notions fully captures the effort disparity. 
\modif{See the Appendix for a numerical example further illustrating why and how our measure of effort unfairness may not be captured by existing notions.}

\begin{figure*}[t!]
	\centering
	\begin{subfigure}[b]{0.3\textwidth}
		\includegraphics[width=\textwidth]{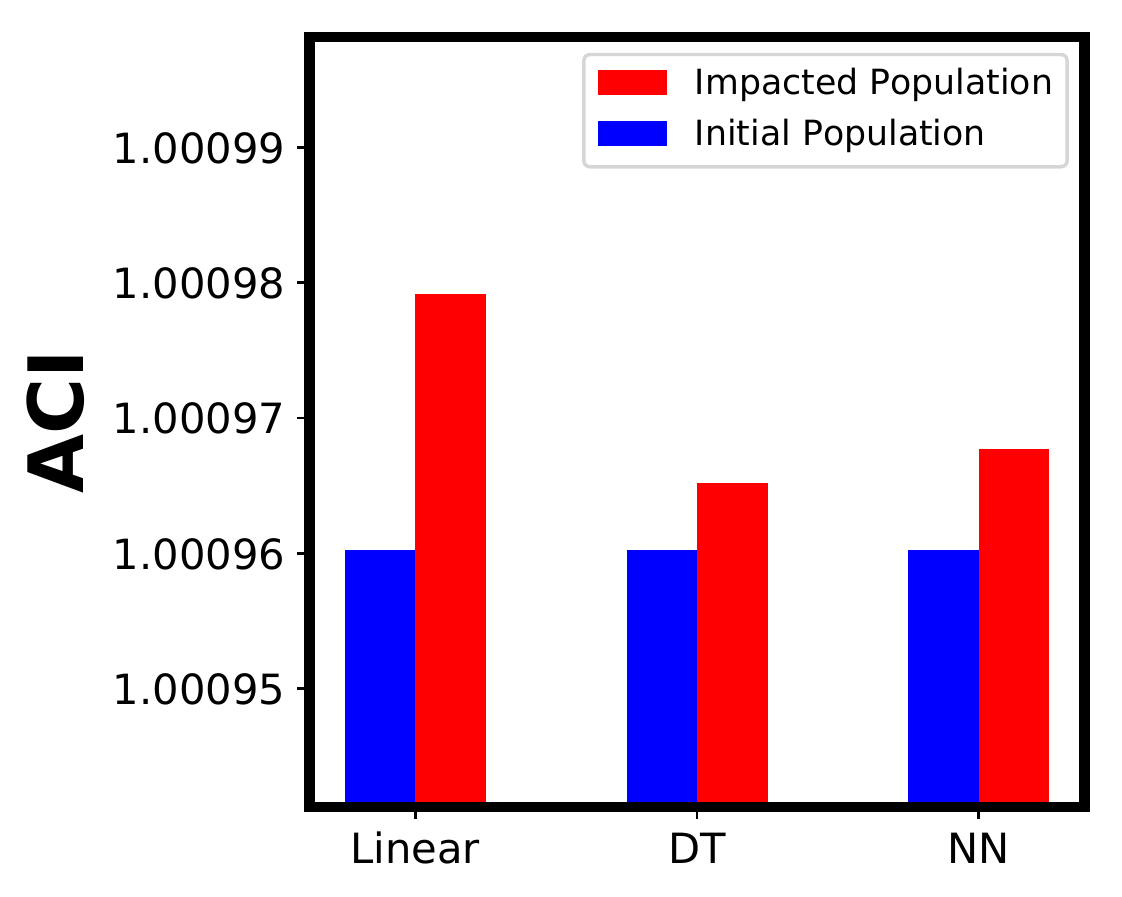}
	\end{subfigure}
	\begin{subfigure}[b]{0.28\textwidth}
		\includegraphics[width=\textwidth]{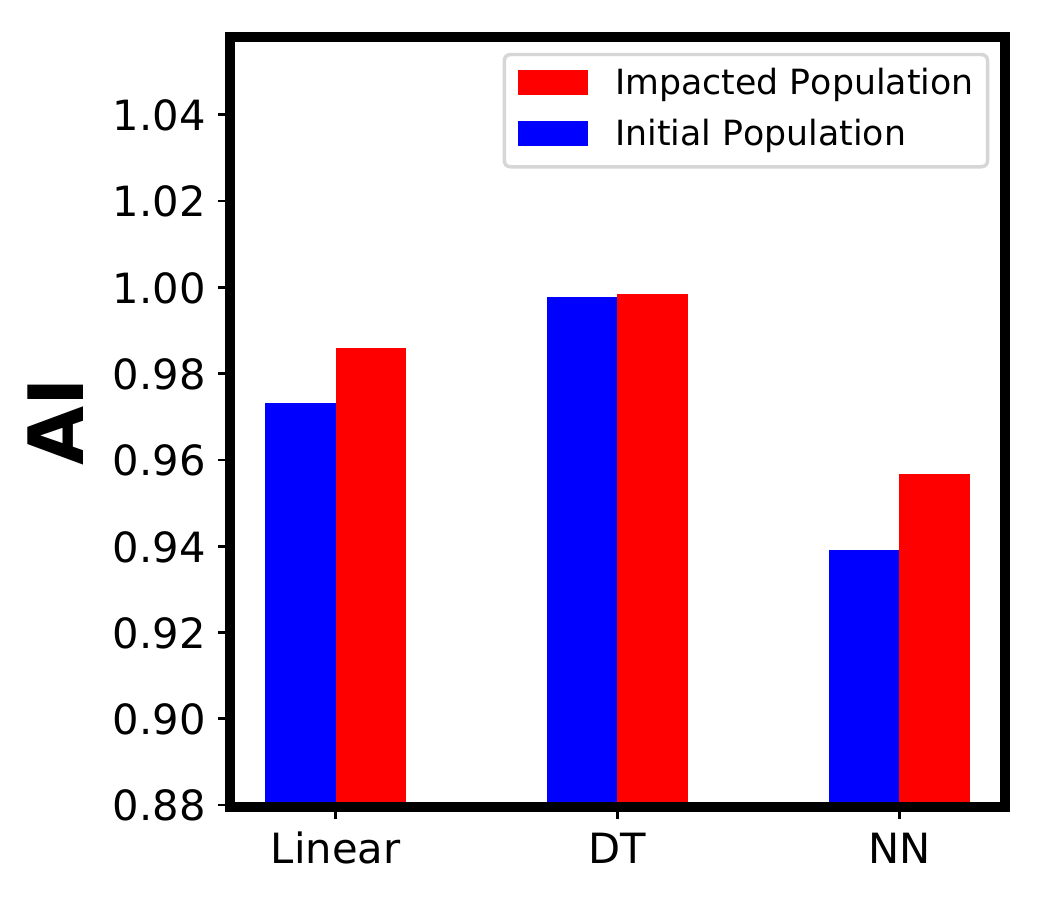}
	\end{subfigure}
	\begin{subfigure}[b]{0.28\textwidth}
		\includegraphics[width=\textwidth]{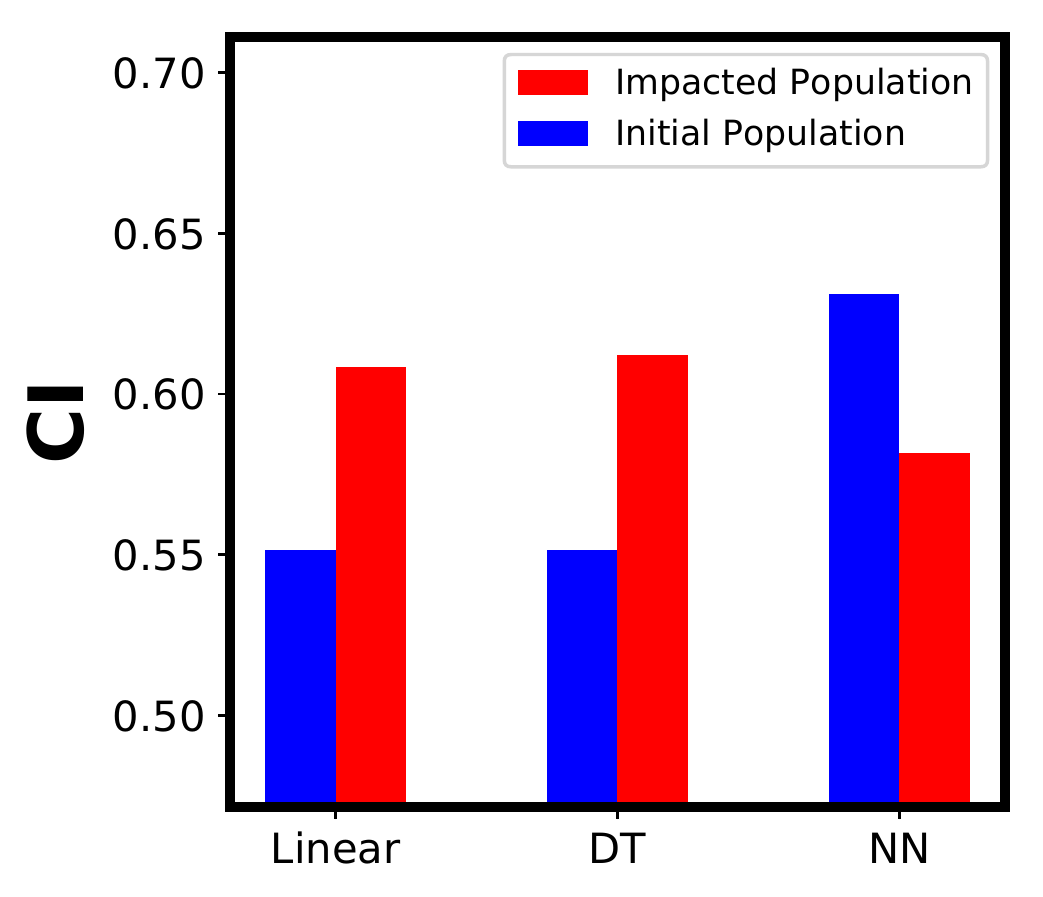}
	\end{subfigure}
	\caption{Segregation measures computed over the initial and impacted population for 3 models. Segregation may increase or decrease as the result of the behavioral dynamics.}\label{fig:segregation}
\end{figure*}


\section{How Algorithmic Policies Re-shape Society}\label{sec:long-term}

To formulate the long-term impact of algorithmic policies on the underlying population, in this section we propose a micro-scale model of how individuals may respond to algorithmic policies, taking inspiration from the psychological literature on \emph{social learning}~\citep{bandura1978social}. We posit that individuals observe and potentially imitate the behavior of their so-called \emph{social models}. A social model is another decision subject who has received a higher level of benefit as the result of being subject to the decision-making model. \modif{Our social learning model captures settings in which subjects don't know the inner workings of the decision making model, but can infer how to improve their standing by observing the decisions it makes for people similar to them (i.e., their social models).}

With the behavioral dynamics specified, we can quantify the macro-scale long-term impact of the model on reshaping the underlying populations. We adapt existing measures of segregation from sociology and economics~\citep{massey1988dimensions} to characterize how the distribution of qualifications for each group changes in response to the deployed model. \modif{Measures of segregation quantify how separate the two subpopulations are in terms of distribution of qualifications. We believe such \emph{model-independent} measures are important to consider, because the decision making model itself may change over time, but its impact on the underlying population may be long lasting.}

\subsection{Behavioral Dynamics}
At a high level, we simulate every individual's response to the predictive model by selecting a social model for them from the training data set; the individual is then assumed to exert effort to attain his/her model's qualifications if and only if doing so improves her overall utility.

Our micro-scale model is meant to capture two important nuances pointed out by the social learning theory:
First, according to the theory observers recreate their social model's behavior only if they have sufficient motivation and this motivation often comes from observing that the social model is rewarded for their actions. We capture this by assuming that an individual recreates his/her social model's qualification if by doing so he/she is sufficiently rewarded and obtains a positive utility.
Second, it has been shown that observers learn best from social models that they identify with and find it within their ability to emulate. These points are captured through our notion of \emph{effort} and \emph{utility}. If a potential social model belongs to a different group than that of the individual, the effort it takes to recreate his/her actions is very high, therefore the individual won't find sufficient utility in imitating him/her.

Assuming that the training data is a representative sample of the population\footnote{We cannot always make this assumption---the sampling process can become biased in numerous ways. }, we select the social models from among the individuals present in the data set. In particular, for an individual $\vz$, the model (denoted by $\vz'$) is another decision subject in $D$ whose imitation would maximize $\vz$'s utility. That is, $\vz' = \arg \max_{\vz'' \in D} \util_h(\vz,\vz'').$ 

\hh{Two remarks are in order. First, note that each one of our fairness notions corresponds to a criterion for choosing the social model. Depending on the context, one criterion may better reflect the human response. In this paper, we deliberately focus on utility maximization. This choice is primarily for ease of illustration---it allows us to forgo specifying $\delta$---but our analysis can be replicated for the other two criteria, as well.} 

Second, one may ask how can individuals be expected to find the right social model (in particular, the utility-maximizing one)? One concrete way is through \emph{actionable} or \emph{counterfactual} explanations~\citep{wachter2017counterfactual,ustun2018actionable}. When an individual fails to receive their desired prediction, such explanations lay out the optimal change he/she can make to improve their outcome. (We re-emphasize that we do not consider our model to be a perfect reflection of extremely nuanced human behavior in the real world. We do, however, consider it to be a reasonable approximation of certain aspects of the process. Our primary goal with this model is to illustrate the role of behavioral dynamics in driving the societal impact of a decision making model.)


Through our proposed dynamics, we can simulate how subjects respond to the predictive model. We then obtain a new data set $D'$ representing the \emph{impacted population}'s qualifications. Next, we adopt measures of segregation to compare the initial and impacted populations and quantify the long-term macro-scale impact of the predictive model on the underlying population. \hh{Our interest in measures of segregation is rooted in the observation that unfairness usually emerges as a concern when there is a clear separation between different segments of the population (in terms of qualifications and/or outcomes). If people belonging to two socially salient groups are fully mixed---in terms of their features and outcomes----such concerns are unlikely to arise. We emphasize that segregation does not always imply unfairness (for instance, segregation could be the consequence of \emph{specialization}: different segments of the population may willingly invest in different sets of qualifications).  But unfairness often comes with some form of segregation. We, therefore, propose measures of segregation as an effective test for potential unfairness. }

\begin{figure*}[t!]
	\centering
	\begin{subfigure}[b]{0.3\textwidth}
		\includegraphics[width=\textwidth]{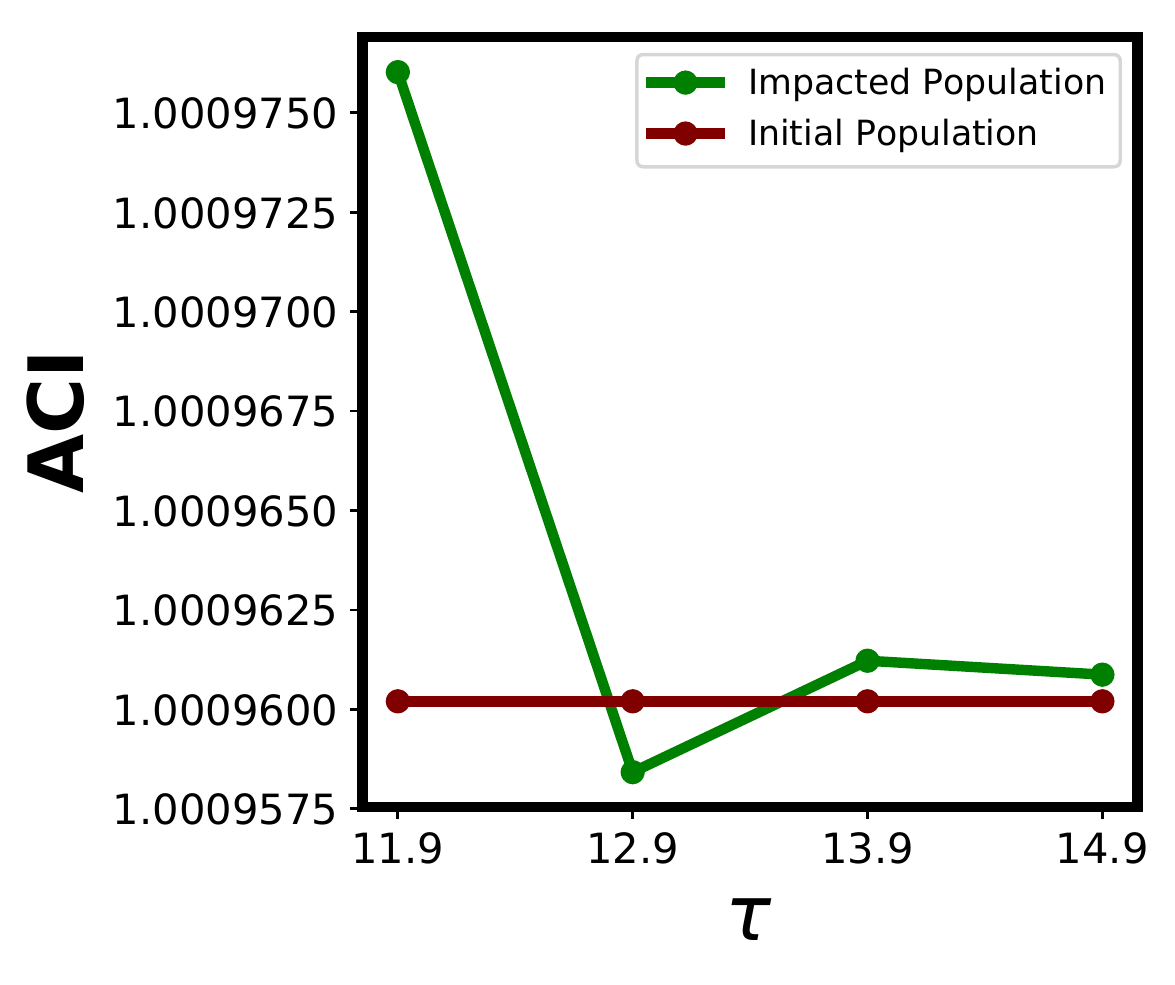}
		\caption{Clustering vs. $\tau$}
		\label{fig:clustering_fc}
	\end{subfigure}
	\begin{subfigure}[b]{0.27\textwidth}
		\includegraphics[width=\textwidth]{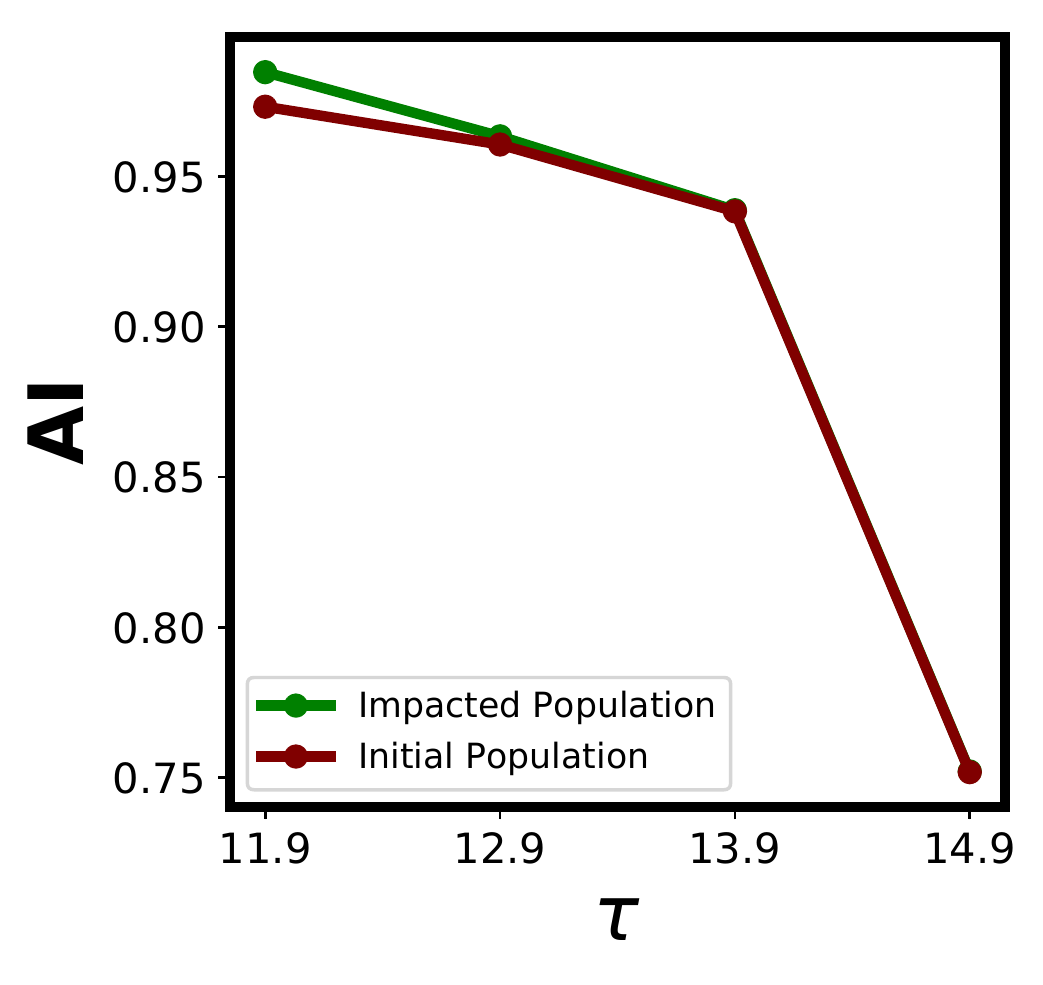}
		\caption{Evenness vs. $\tau$}
		\label{fig:evenness_fc}
	\end{subfigure}
		\begin{subfigure}[b]{0.27\textwidth}
		\includegraphics[width=\textwidth]{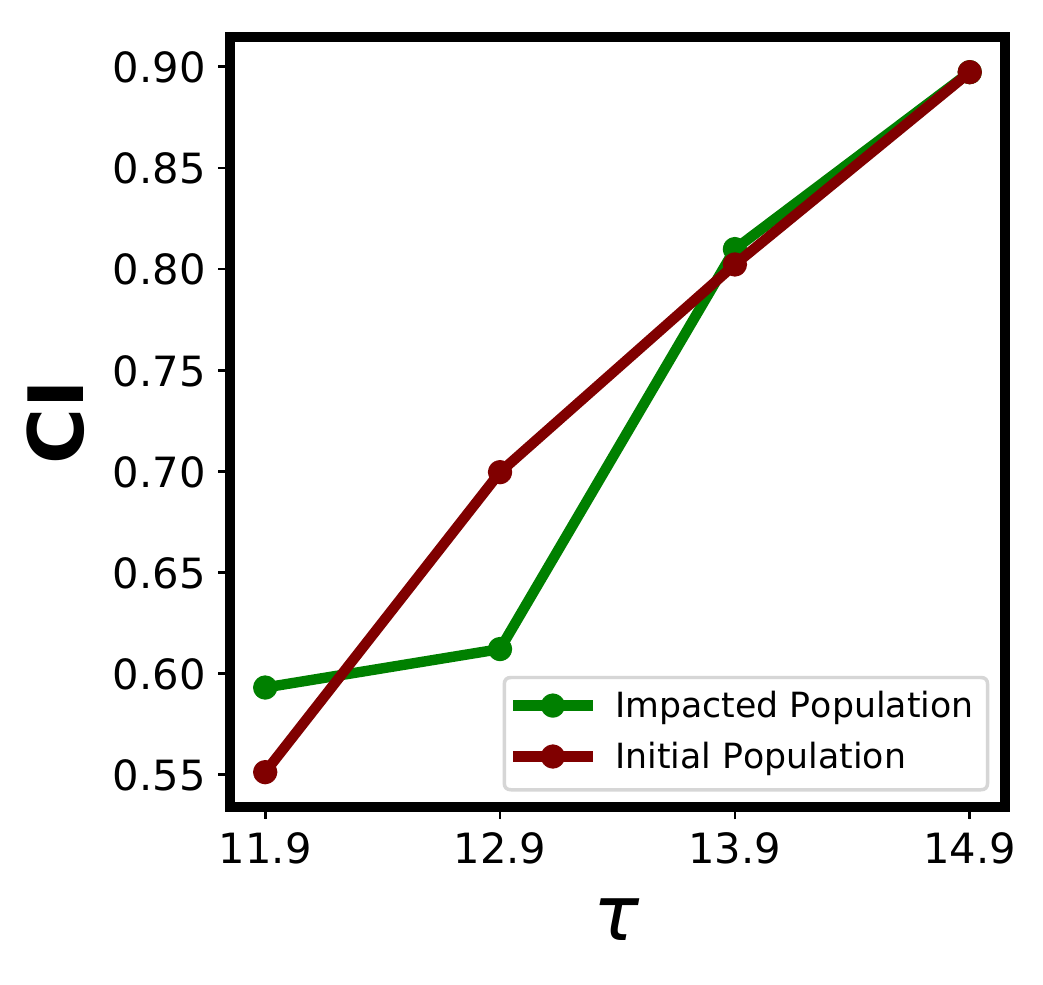}
		\caption{Centralization vs. $\tau$}
		\label{fig:centralization_fc}
	\end{subfigure}
	\caption{The change in various measures of segregation as a function of enforcing fairness contraints with strength $\tau$.}\label{fig:constraints}
\end{figure*}

\subsection{Quantifying Impact via Measures of Segregation}\label{sec:segregation}
Ethnic and racial segregation is a well-studied phenomenon in sociology. At a high level,  segregation is the degree to which two or more groups live separately from one another. A long line of work in sociology has been concerned with measuring segregation. In their highly influential article, \citet{massey1988dimensions} break down residential segregation into five distinct axes of measurement: centralization, evenness, clustering, exposure, and concentration. Below, we overview these measures and show how three of them can be utilized to measure the macro-scale impact of decision-making models on the distribution of qualifications.

\textbf{Evenness} measures how unevenly the minority group is distributed over areal units. Evenness is maximized when all units have the same relative number of minority and majority members as the whole population.
More precisely, for an area/neighborhood $i$, let $t_i$ denote its total population, $m_i$ the number of minority residents, and $M_i$ the number of majority residents of the neighborhood. Also, let $p_i = m_i/t_i$ specify the percentage of minority residents in the area. Let $T$ and $P$ specify the total population size and minority proportion of the whole population. Suppose there are $N$ areal units in total. 
The Atkinson Index (AI) is a particular measure of evenness satisfying several desirable properties.\footnote{It satisfies the \textit{transfer principle}, \textit{compositional invariance}, \textit{population invariance}, and \textit{organizational equivalence}. 
}
 For a constant $0 < \beta < 1$, the Atkinson index measures the inequality of $(1-p_i)/p_i$ (i.e. the number of majority residents per minority resident in neighborhood $i$) computed across all individuals belonging to the minority group:
$
1-\frac{P}{1-P}\left(\frac{1}{N}\sum_{i=1}^{N} (1-p_i)^{1-\beta} p_i^\beta t_i / TP\right)^{1/(1-\beta)}.
$

\textbf{Centralization} is the degree to which a group is spatially located near the center of an urban area. (Because of certain urban development policies in the past, central areas of most cities across the U.S. are declining residential areas.) The degree of centralization can be measured by comparing the percentage of minority residents living in the central areas of the city. The Centralization Index (CI) is precisely defined as follows: $\frac{\sum_{i \text{ central}} m_i}{m}$ where $m$ is the total minority population.

\textbf{Clustering} measures the extent to which areal units inhabited by minority members adjoin one another, or cluster, in space. For example, the \emph{Absolute Clustering Index} (ACI) ``expresses the average number of [minority] members in nearby [areal units] as a proportion of the total population in those nearby [areal units]"~\citep{massey1988dimensions}. ACI is defined as follows:
\begin{equation*}\label{eq:ACI}
\frac{ \left[ \sum_{i=1}^N \sum_{j=1}^N c_{i,j} \frac{m_i}{m} m_j \right]  - \left[ \sum_{i=1}^N \sum_{j=1}^N c_{i,j} \frac{1}{N} \frac{m}{N} \right] }{ \left[ \sum_{i=1}^N \sum_{j=1}^N c_{i,j} \frac{m_i}{m} t_j \right]  - \left[ \sum_{i=1}^N \sum_{j=1}^N c_{i,j} \frac{1}{N} \frac{m}{N} \right]}.
\end{equation*}
For any two areas $i$ and $j$, $c_{i,j}$ specifies the closeness between their corresponding centers. 

Residential \textbf{exposure} refers to the degree of potential contact, or the possibility of interaction, between minority and majority group members within geographic areas of a city. 
\textbf{Concentration} refers to relative amount of physical space occupied by a minority group in the urban environment. 



There are several notable differences between our setting and that of residential segregation. First, in our setting there are no predefined notions of ``area" or ``neighborhoods" that individuals belong to. Second, individuals are described by multi-dimentional feature vectors---as opposed to a 2-dimensional vector specifying their residential location. Third, it is not immediately clear how distances and similarity between individuals should be defined. 
Next, we will address these issues for evenness, centralization, and clustering.

Throughout this section, we will focus on the mutable feature subspace.
We take the distance between two individuals $i$ (belonging to group $s$) and $j$ (belonging to group $s'$) as follows: $d(i,j) = \max\{d^s(i , j), d^{s'}(i , j)\}$, where $d^s(i , j)$ is defined as follows: $\max\{ 0, Q_s(y_j) - Q_s(y_i) \} + \sum_{k \in \text{ mutable}} \epsilon_{s,k}(x_k,x'_k)$.
We will use Atkinson index to measure evenness. We will specify areas through what we call \emph{focal points}---these are feature vectors in the mutable feature subspace that at least one subject in the original population imitates to improve their utility. Each focal point corresponds to a neighborhood, and an individual belongs to the neighborhood of their nearest focal point. 
We measure the degree of centralization by comparing the percentage of minority individuals whose predictions are above the average (e.g., $\hY > 11.94$).
We measure ACI at the individual level---that is, we assume each individual corresponds to a neighborhood. We define the similarity between two neighborhoods as follows: $c_{i,j} = e^{-d(i,j)}$. 

Figure~\ref{fig:segregation} illustrates our measures of segregation both for the initial population (depicted in blue) and the impacted population---after individuals respond to the model by adjusting their qualifications (depicted in red). Segregation can change in counter-intuitive ways through imitation dynamics. 






\section{The Role of Fairness Interventions}\label{sec:mechanism}
In this Section, we investigate the effect of enforcing fairness constraints---at the time of training---on the long term population-level impact of the deployed model. We focus on the case of linear regression. We train a model by minimizing the mean squared error while imposing the welfare constraints proposed by \citet{heidari2018fairness}. 

Figure~\ref{fig:constraints} shows the effect of imposing fairness constraints on various measures of segregation (all computed by taking females as the minority/protected group).
One might expect that these constraints would always reduce segregation in the long run. As illustrated in Figure~\ref{fig:constraints}, this is not always the case. 
For a small value of $\tau$, enforcing fairness constraints can significantly reduce the degree of clustering (see Figure~\ref{fig:clustering_fc}). Larger values of $\tau$ can reverse this effect and lead to a population that is more heavily clustered/segregated compared to the original population. 
Evenness remains relatively unchanged regardless of the value of $\tau$ (see Figure~\ref{fig:evenness_fc}).

These findings highlight an important insight about fairness constraints: they can affect segregation in two competing ways. On the one hand by automatically assigning a desirable label to some members of the disadvantaged group, the model incentivizes these members not to make any change. On the other hand, these members can serve as social models for the rest of the disadvantaged group, nudging more of them to improve their qualifications and obtain better labels. Which force is more powerful? One can only answer this by simulating the dynamics on the particular data set at hand. 
\modif{We see clear parallels between our observations and the prior work on affirmative action policies. Advocates of affirmative action often argue that a larger representation of minorities in desirable positions can lead to role models who encourage other minorities in their investment decisions (see e.g., \citep{chung2000role}). At the same time, critics argue that affirmative action quota may indirectly harm the disadvantaged group members by reducing their incentives to invest in qualifications~\citep{coate1993antidiscrimination,coate1993will}. Similar to our work, economic results on the long-term impact of affirmative action policies is mixed and context-specific. } 

\hh{
We end this section with a remark on fairness-restoring interventions. While we focused on \emph{algorithmic} interventions, we must emphasize that changing the decision-making model is not the only mechanism through which segregation and unfairness can be alleviated. Instead of artificially changing the decision boundary, it may be socially more desirable to address unfairness \emph{before} people are subjected to algorithmic decision making. For instance, one could design and implement policies that make it easier for disadvantaged group members to obtain certain qualifications. We leave the analysis of such \emph{feature interventions} as a promising direction for future work.
}

\section{Conclusion \& Future Directions}\label{sec:future}

We presented a data-driven framework for studying the potential long-term impact of predictive models on decision subjects and society. We proposed a micro-model of human response to algorithmic policies rooted in psychology and several macro-level measures of change borrowed from sociology and economics. 
Our work suggests several immediate directions for future work, including but not limited to (a) human subject experiments to investigate the viability of our behavioral model; (b) designing an efficient mechanism for bounding effort-reward unfairness.

\section*{Acknowledgements}
K. P. Gummadi is supported in part by an ERC Advanced Grant ``Foundations for Fair Social Computing'' (no. 789373).

\bibliography{MyBib}

\begin{thebibliography}{38}
\providecommand{\natexlab}[1]{#1}
\providecommand{\url}[1]{\texttt{#1}}
\expandafter\ifx\csname urlstyle\endcsname\relax
  \providecommand{\doi}[1]{doi: #1}\else
  \providecommand{\doi}{doi: \begingroup \urlstyle{rm}\Url}\fi

\bibitem[Angwin et~al.(2016)Angwin, Larson, Mattu, and Kirchner]{propublica}
Angwin, J., Larson, J., Mattu, S., and Kirchner, L.
\newblock Machine bias.
\newblock \emph{Propublica}, 2016.

\bibitem[Apesteguia et~al.(2007)Apesteguia, Huck, and
  Oechssler]{apesteguia2007imitation}
Apesteguia, J., Huck, S., and Oechssler, J.
\newblock Imitation---theory and experimental evidence.
\newblock \emph{Journal of Economic Theory}, 136\penalty0 (1):\penalty0
  217--235, 2007.

\bibitem[Bandura(1962)]{bandura1962social}
Bandura, A.
\newblock Social learning through imitation.
\newblock 1962.

\bibitem[Bandura(1978)]{bandura1978social}
Bandura, A.
\newblock Social learning theory of aggression.
\newblock \emph{Journal of communication}, 28\penalty0 (3):\penalty0 12--29,
  1978.

\bibitem[Bandura(2008)]{bandura2008observational}
Bandura, A.
\newblock Observational learning.
\newblock \emph{The international encyclopedia of communication}, 2008.

\bibitem[Barry-Jester et~al.(2015)Barry-Jester, Casselman, and
  Goldstein]{sentencing}
Barry-Jester, A., Casselman, B., and Goldstein, D.
\newblock The new science of sentencing.
\newblock \emph{The Marshall Project}, August 2015.

\bibitem[Calders et~al.(2013)Calders, Karim, Kamiran, Ali, and
  Zhang]{calders2013controlling}
Calders, T., Karim, A., Kamiran, F., Ali, W., and Zhang, X.
\newblock Controlling attribute effect in linear regression.
\newblock In \emph{Proceedings of the International Conference on Data Mining},
  pp.\  71--80. IEEE, 2013.

\bibitem[Chung(2000)]{chung2000role}
Chung, K.-S.
\newblock Role models and arguments for affirmative action.
\newblock \emph{American Economic Review}, 90\penalty0 (3):\penalty0 640--648,
  2000.

\bibitem[Coate \& Loury(1993{\natexlab{a}})Coate and
  Loury]{coate1993antidiscrimination}
Coate, S. and Loury, G.
\newblock Antidiscrimination enforcement and the problem of patronization.
\newblock \emph{The American Economic Review}, 83\penalty0 (2):\penalty0
  92--98, 1993{\natexlab{a}}.

\bibitem[Coate \& Loury(1993{\natexlab{b}})Coate and Loury]{coate1993will}
Coate, S. and Loury, G.~C.
\newblock Will affirmative-action policies eliminate negative stereotypes?
\newblock \emph{The American Economic Review}, pp.\  1220--1240,
  1993{\natexlab{b}}.

\bibitem[Corbett-Davies et~al.(2017)Corbett-Davies, Pierson, Feller, Goel, and
  Huq]{corbett2017algorithmic}
Corbett-Davies, S., Pierson, E., Feller, A., Goel, S., and Huq, A.
\newblock Algorithmic decision making and the cost of fairness.
\newblock In \emph{Proceedings of the 23rd ACM SIGKDD International Conference
  on Knowledge Discovery and Data Mining}, pp.\  797--806. ACM, 2017.

\bibitem[Cortez \& Silva(2008)Cortez and Silva]{cortez2008using}
Cortez, P. and Silva, A. M.~G.
\newblock Using data mining to predict secondary school student performance.
\newblock 2008.

\bibitem[Deo(2015)]{deo2015machine}
Deo, R.~C.
\newblock Machine learning in medicine.
\newblock \emph{Circulation}, 132\penalty0 (20):\penalty0 1920--1930, 2015.

\bibitem[Dong et~al.(2018)Dong, Roth, Schutzman, Waggoner, and
  Wu]{dong2018strategic}
Dong, J., Roth, A., Schutzman, Z., Waggoner, B., and Wu, Z.~S.
\newblock Strategic classification from revealed preferences.
\newblock In \emph{Proceedings of the 2018 ACM Conference on Economics and
  Computation}, pp.\  55--70. ACM, 2018.

\bibitem[Duivesteijn \& Feelders(2008)Duivesteijn and
  Feelders]{duivesteijn2008nearest}
Duivesteijn, W. and Feelders, A.
\newblock Nearest neighbour classification with monotonicity constraints.
\newblock In \emph{Joint European Conference on Machine Learning and Knowledge
  Discovery in Databases}, pp.\  301--316. Springer, 2008.

\bibitem[Dwork et~al.(2012)Dwork, Hardt, Pitassi, Reingold, and
  Zemel]{dwork2012fairness}
Dwork, C., Hardt, M., Pitassi, T., Reingold, O., and Zemel, R.
\newblock Fairness through awareness.
\newblock In \emph{Proceedings of the Innovations in Theoretical Computer
  Science Conference}, pp.\  214--226. ACM, 2012.

\bibitem[Feldman et~al.(2015)Feldman, Friedler, Moeller, Scheidegger, and
  Venkatasubramanian]{feldman2015certifying}
Feldman, M., Friedler, S.~A., Moeller, J., Scheidegger, C., and
  Venkatasubramanian, S.
\newblock Certifying and removing disparate impact.
\newblock In \emph{Proceedings of the International Conference on Knowledge
  Discovery and Data Mining}, pp.\  259--268. ACM, 2015.

\bibitem[Hardt et~al.(2016)Hardt, Price, and Srebro]{hardt2016equality}
Hardt, M., Price, E., and Srebro, N.
\newblock Equality of opportunity in supervised learning.
\newblock In \emph{Proceedings of the 30th Conference on Neural Information
  Processing Systems}, pp.\  3315--3323, 2016.

\bibitem[Heidari et~al.(2018)Heidari, Ferrari, Gummadi, and
  Krause]{heidari2018fairness}
Heidari, H., Ferrari, C., Gummadi, K.~P., and Krause, A.
\newblock Fairness behind a veil of ignorance: A welfare analysis for automated
  decision making.
\newblock In \emph{Proceedings of the 32nd Conference on Neural Information
  Processing Systems}, 2018.

\bibitem[Heidari et~al.(2019)Heidari, Loi, Gummadi, and Krause]{heidari2019a}
Heidari, H., Loi, M., Gummadi, K.~P., and Krause, A.
\newblock A moral framework for understanding of fair ml through economic
  models of equality of opportunity.
\newblock In \emph{Proceedings of the 2nd ACM Conference on Fairness,
  Accountability, and Transparency}, 2019.

\bibitem[Hu et~al.(2019)Hu, Immorlica, and Vaughan]{hu2018disparate}
Hu, L., Immorlica, N., and Vaughan, J.~W.
\newblock The disparate effects of strategic manipulation.
\newblock In \emph{Proceedings of the 2nd ACM Conference on Fairness,
  Accountability, and Transparency}, 2019.

\bibitem[Kannan et~al.(2019)Kannan, Roth, and Ziani]{kannan2018downstream}
Kannan, S., Roth, A., and Ziani, J.
\newblock Downstream effects of affirmative action.
\newblock In \emph{Proceedings of the 2nd ACM Conference on Fairness,
  Accountability, and Transparency}, 2019.

\bibitem[Kleinberg et~al.(2017)Kleinberg, Mullainathan, and
  Raghavan]{kleinberg2016inherent}
Kleinberg, J., Mullainathan, S., and Raghavan, M.
\newblock Inherent trade-offs in the fair determination of risk scores.
\newblock In \emph{In proceedings of the 8th Innovations in Theoretical
  Computer Science Conference}, 2017.

\bibitem[Levin(2016)]{guardian_beauty}
Levin, S.
\newblock A beauty contest was judged by {AI} and the robots didn't like dark
  skin.
\newblock \emph{The Guardian}, 2016.

\bibitem[Liu et~al.(2018)Liu, Dean, Rolf, Simchowitz, and
  Hardt]{liu2018delayed}
Liu, L.~T., Dean, S., Rolf, E., Simchowitz, M., and Hardt, M.
\newblock Delayed impact of fair machine learning.
\newblock In \emph{Proceedings of the International Coference on Machine
  Learning}, 2018.

\bibitem[Massey \& Denton(1988)Massey and Denton]{massey1988dimensions}
Massey, D.~S. and Denton, N.~A.
\newblock The dimensions of residential segregation.
\newblock \emph{Social forces}, 67\penalty0 (2):\penalty0 281--315, 1988.

\bibitem[Milli et~al.(2019)Milli, Miller, Dragan, and Hardt]{milli2018social}
Milli, S., Miller, J., Dragan, A.~D., and Hardt, M.
\newblock The social cost of strategic classification.
\newblock In \emph{Proceedings of the 2nd ACM Conference on Fairness,
  Accountability, and Transparency}, 2019.

\bibitem[Mullainathan \& Thaler(2000)Mullainathan and
  Thaler]{mullainathan2000behavioral}
Mullainathan, S. and Thaler, R.~H.
\newblock Behavioral economics.
\newblock Technical report, National Bureau of Economic Research, 2000.

\bibitem[Petrasic et~al.(2017)Petrasic, Saul, Greig, and Bornfreund]{whitecase}
Petrasic, K., Saul, B., Greig, J., and Bornfreund, M.
\newblock Algorithms and bias: What lenders need to know.
\newblock \emph{White \& Case}, 2017.

\bibitem[Roemer \& Trannoy(2015)Roemer and Trannoy]{roemer2015equality}
Roemer, J.~E. and Trannoy, A.
\newblock Equality of opportunity.
\newblock In \emph{Handbook of income distribution}, volume~2, pp.\  217--300.
  Elsevier, 2015.

\bibitem[Rudin(2013)]{policing}
Rudin, C.
\newblock Predictive policing using machine learning to detect patterns of
  crime.
\newblock \emph{Wired Magazine}, August 2013.
\newblock Retrieved 4/28/2016.

\bibitem[Sandholm(2010)]{sandholm2010population}
Sandholm, W.~H.
\newblock \emph{Population games and evolutionary dynamics}.
\newblock MIT press, 2010.

\bibitem[Sen(1993)]{sen1993capability}
Sen, A.
\newblock Capability and well-being.
\newblock \emph{The quality of life}, 30, 1993.

\bibitem[Sweeney(2013)]{sweeney2013discrimination}
Sweeney, L.
\newblock Discrimination in online ad delivery.
\newblock \emph{Queue}, 11\penalty0 (3):\penalty0 10, 2013.

\bibitem[Tajfel et~al.(1979)Tajfel, Turner, Austin, and
  Worchel]{tajfel1979integrative}
Tajfel, H., Turner, J.~C., Austin, W.~G., and Worchel, S.
\newblock An integrative theory of intergroup conflict.
\newblock \emph{Organizational identity: A reader}, pp.\  56--65, 1979.

\bibitem[Ustun et~al.(2018)Ustun, Spangher, and Liu]{ustun2018actionable}
Ustun, B., Spangher, A., and Liu, Y.
\newblock Actionable recourse in linear classification.
\newblock In \emph{Proceedings of the 2nd ACM Conference on Fairness,
  Accountability, and Transparency}, 2018.

\bibitem[Wachter et~al.(2017)Wachter, Mittelstadt, and
  Russell]{wachter2017counterfactual}
Wachter, S., Mittelstadt, B., and Russell, C.
\newblock Counterfactual explanations without opening the black box: Automated
  decisions and the {GDPR}.
\newblock 2017.

\bibitem[Zafar et~al.(2017)Zafar, Valera, Gomez~Rodriguez, and
  Gummadi]{zafar2017fairness}
Zafar, M.~B., Valera, I., Gomez~Rodriguez, M., and Gummadi, K.~P.
\newblock Fairness constraints: Mechanisms for fair classification.
\newblock In \emph{Proceedings of the 20th International Conference on
  Artificial Intelligence and Statistics}, 2017.

\end{thebibliography}
\bibliographystyle{icml2019}

 \appendix

\begin{figure*}[t!]
	\raggedleft
	\begin{subfigure}{\textwidth}
	\includegraphics[width=1\textwidth]{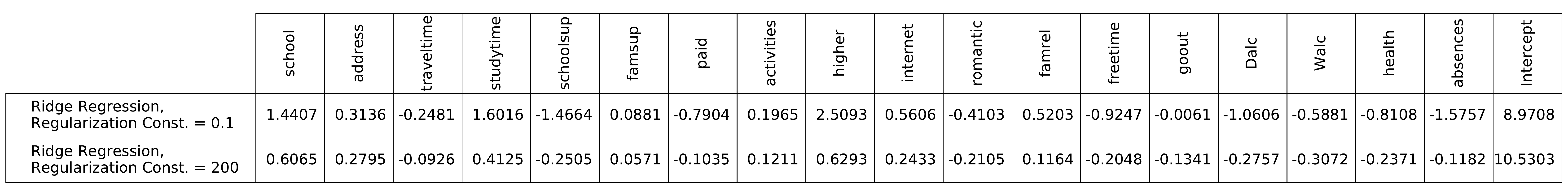}
	\caption{Mutable feature weights (and intercept) common to both models.}
	\label{fig:common}
	\end{subfigure}

	\begin{subfigure}{\textwidth}
	\includegraphics[width=1\textwidth]{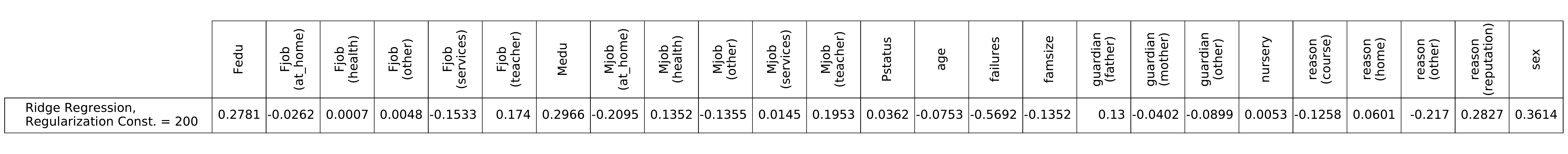}
	\caption{Immutable feature weights; only applicable to Ridge Regression with regularization const = 200.}
	\label{fig:immutable}
	\end{subfigure}

	\caption{Weights assigned by 2 models, one trained with only mutable features (top row in Figure~\ref{fig:common}) and the other trained with both mutable and immutable features (bottom row in Figure~\ref{fig:common} and the model in Figure~\ref{fig:immutable}).}
	\label{fig:all_features_regression_example}
\end{figure*}

\section{The Effort Function}
We define $\epsilon_{s,k}(x_k,x'_k,)$---depending on the type of feature $k$---as follows:
\begin{itemize}
	\item \textbf{Non-monotone numerical feature:} Suppose feature $k$ is numerical, but it is not clear which direction of change should increase the probability of the instance $\vx$ being labeled as positive. An example of this type of feature in the education context is \emph{extracurricular activities}---depending on other factors this may be increase or decrease one's performance in school.  
	For this type of feature, we assume change in either direction requires effort, and define $\epsilon_{s,k}(x_k,x'_k,)$ as follows:
	$$\epsilon_{s,k}(x_k,x'_k) = \vert Q_{s,k}(x'_k) - Q_{s,k}(x_k) \vert.$$
	%
	
	\item \textbf{Ordinal feature:} We define $\epsilon_{s,k}(x_k,x'_k,)$ similar to numerical features---depending on whether we consider the attribute monotone or not. 
	\item \textbf{Categorical feature:} Suppose feature $k$ is categorical and can take on $n_k$ different values $\{ v_1, \cdots, v_{n_k} \}$ (example: marital status). We define $\epsilon_{s,k}$ via $n^2_K$ constants, $c_{i,j}$ for $1 \leq i,j \leq n_k$, with $c_{i,j}$ specifying the effort required to change the value of feature $k$ from $v_i$ to $v_j$. 
	Throughout our simulations and for simplicity, we assume there exists a constant $c$ such that $c_{i,j} \equiv c$ for all $1 \leq i,j \leq n_k$. 
	\item \textbf{(Conditionally) immutable feature:} We call feature $k$ \emph{(conditionally) immutable} if there exist two values $x_k \neq x'_k$, where the change from $x_k$ to $x'_k$ is considered impossible. For example, race is an immutable feature (one cannot be expected to change their race). Age is conditionally immutable (one cannot be expected to become younger). In this case we define our effort function as follows: $\epsilon_{s,k}(x_k,x'_k) =  \infty.$
\end{itemize}

\section{The Student Performance Data Set}\label{sec:student_dataset}
The student performance data set~\citep{cortez2008using} contains information about student achievement in secondary education of two Portuguese schools. The data attributes include student grades, demographic, social and school related features. 
The data set consists of 649 instances/student, with each instance consisting of 32 features. The task is to predict the student's final grade (value from 0 to 20) in Portuguese. Out of the 32 features, we choose only features that are considered mutable in at least one direction, that is, the student can exert effort and change the feature value. We dropped all immutable features---except gender---to be able to find a social model for every student. (Since the data set is very small, this would not have been possible had we kept the immutable features). This results in a total of 23 features out of which 10 are binary and the rest are numerical. 
We then perform a 70:30 train-test split, with the train set consisting of 454 instances and the test set consisting of 195 instances.

\section{Trained Models}
We trained the following models on the student performance data set:
\begin{itemize}
\item \textbf{Neural network:} A shallow neural network with one hidden layer (ReLu activation) containing 100 nodes. Loss function with L2 regularization with regularization strength = 10. Regularization strength and number of nodes in the hidden layer were found using grid search by doing a 3 fold cross validation and taking the parameters that resulted in the maximum average test accuracy. 
\item \textbf{Linear regressor:} Least squares solver. Finds parameters B such that L2 norm of $|Bx - Y|$ is minimized.
\item \textbf{Decision Tree:} Decision Tree Regressor with maximum depth of 5 to avoid overfitting. Max depth parameter was chosen using grid search by doing a 3-fold cross validation and choosing the parameter that maximised the average test set accuracy. Criterion for splitting was minimization of MSE.
\end{itemize}

\section{Fairness Notions for Regression}\label{sec:fairness_notions}
\textbf{Positive residual difference}~\citep{calders2013controlling} is computed by taking the absolute difference of mean positive residuals across groups:
{\scriptsize
$$\left|\frac{1}{|G_1^+|} \sum_{i \in G_1} \max\{0,(\hat{y}_i- y_i)\}  - \frac{1}{|G_2^+|}\sum_{i \in G_2} \max\{0,(\hat{y}_i -y_i)\}\right|.$$
}
\textbf{Negative residual difference}~\citep{calders2013controlling} is computed by taking the absolute difference of mean negative residuals across groups:
{\scriptsize
$$\left|\frac{1}{|G_1^-|} \sum_{i \in G_1} \max\{0,(y_i - \hat{y}_i)\}  - \frac{1}{|G_2^-|}\sum_{i \in G_2} \max\{0,(y_i - \hat{y}_i)\}\right|.$$
}

%

\section{Why Existing Notions of Fairness Fail to Capture Effort-Reward Disparity}

Figure~\ref{fig:all_features_regression_example} shows an example of 2 ridge regressions, both trained on the student performance dataset (described in section~\ref{sec:student_dataset}), but one has access to only mutable features and the other has access to both mutable and (conditionally) immutable features. For simplicity, let's call them ``mutable model'' and ``combined model'' respectively. Both the ``mutable model'' and ``combined model'' have similar error distributions on the dataset with Mean Averaged Errors (MAE) of $2.028$ and $2.046$ on the entire population. They also have similar errors across sub-groups defined based on the value of sensitive feature $s$ (for the student dataset, $s$ corresponds to gender); with MAEs for the sub-group with $s = 1$ (females) being $1.999$ and $2.067$ and MAEs for sub-group with $s = 0$ (males) being $2.068$ and $2.016$ for ``mutable model'' and ``combined model'' respectively. Lastly, both the models also have comparable measures of existing fairness notions defined in section~\ref{sec:fairness_notions} with positive residuals of $0.296$ and $0.228$ and negative residuls of $0.237$ and $0.249$ respectively.

However, when evaluated for the effort-reward unfairness, ``mutable model'' and ``combined model'' perform differently with measures of $0.043$ and $0.532$ respectively. One of the reasons for such contrasting values is the different weights each model assigns to the mutable features (shown in Figure~\ref{fig:common}). For example, consider a student at a benefit level of $b_{initial}$ (assuming benefit function = predicted value by the model) subject to predictions by the ``mutable model'' (top row in Figure ~\ref{fig:common}), were to imitate a role model having value of the continuous feature, ``studytime'', greater by 1 unit. Assume, for simplicity, that all other feature values of the role model and the student are same. Say the effort exerted to make this change is $e$ which brings the student to a benefit level of $b$ (=new predicted value by the model), thus making utility, $u_{mutable}$ = $b - b_{initial} - e$. Now say the same student were subject to predictions by the ``combined model'' (bottom row in Figure ~\ref{fig:common}) and were to immitate the same role model (having ``studytime'' greater by 1 unit and having all other features same as the student) as in the previous case. Since both the models have similar prediction errors, we can assume that the student has a similar prediction value as in the previous case (thus being at the same benefit level of $b_{initial}$). The effort is independent of the model, so effort in this case remains $e$. However, since the weight assigned by ``combined model'' to ``studytime'' is $~0.25$x the weight assigned by ``mutable model'' (see Figure~\ref{fig:common}), increasing ``studytime'' by 1 unit will result in a new benefit level of $b^{'} (< b)$. Thus utility in this case, $u_{combined}$ = $b^{'} - b_{initial} - e$. Since $b_{initial}$, $b$, $b^{'}$ and $e$ are all positive values, $u_{mutable} > u_{combined}$ . Thus, the values of utility can differ considerably for 2 models even though their error distributions across the population may be very similar. Our notion of effor-reward unfairness captures this disparity while existing notions of fairness might not.


\end{document}